\documentclass[conference]{IEEEtran}

\usepackage{times}
\usepackage{helvet}
\usepackage{courier}
\usepackage{url}
\usepackage{graphicx}
\usepackage{amsmath}
\usepackage{amsfonts}
\usepackage{color}
\usepackage{booktabs}

\begin{document}

\title{Collaborative Filtering with Graph-based Implicit Feedback}
\author{\IEEEauthorblockN{Minzhe~Niu}
\IEEEauthorblockA{Shanghai Jiao Tong University\\ Shanghai, China\\ nmzfrank@apex.sjtu.edu.cn}
\and
\IEEEauthorblockN{Weinan~Zhang}
\IEEEauthorblockA{Shanghai Jiao Tong University\\ Shanghai, China\\ wnzhang@sjtu.edu.cn}
\and
\IEEEauthorblockN{Yanru~Qu}
\IEEEauthorblockA{Shanghai Jiao Tong University\\ Shanghai, China\\ kevinqu@apex.sjtu.edu.cn}
\and
\IEEEauthorblockN{Xuezhi~Cao}
\IEEEauthorblockA{Shanghai Jiao Tong University\\ Shanghai, China\\ cxz@apex.sjtu.edu.cn}
\and
\IEEEauthorblockN{Ruiming~Tang}
\IEEEauthorblockA{Huawei Noah's Ark Lab\\ Shenzhen, China\\ tangruiming@huawei.com}
\and
\IEEEauthorblockN{Xiuqiang~He}
\IEEEauthorblockA{Huawei Noah's Ark Lab\\ Shenzhen, China\\ hexiuqiang@huawei.com}
\and
\IEEEauthorblockN{Yong~Yu}
\IEEEauthorblockA{Shanghai Jiao Tong University\\ Shanghai, China\\ yyu@apex.sjtu.edu.cn}
}

\maketitle

\begin{abstract}
Introducing consumed items as users' implicit feedback in matrix factorization (MF) method, SVD++ is one of the most effective collaborative filtering methods for personalized recommender systems. 
Though powerful, SVD++ has two limitations: 
(i). only user-side implicit feedback is utilized, whereas item-side implicit feedback, which can also enrich item representations, is not leveraged;
(ii). in SVD++, the interacted items are equally weighted when combining the implicit feedback, which can not reflect user's true preferences accurately.
To tackle the above limitations, in this paper we propose Graph-based collaborative filtering (GCF) model, Weighted Graph-based collaborative filtering (W-GCF) model and Attentive Graph-based collaborative filtering (A-GCF) model, which (i). generalize the implicit feedback to item side based on the user-item bipartite graph; (ii). flexibly learn the weights of individuals in the implicit feedback hence improve the model's capacity.
Comprehensive experiments show that our proposed models outperform state-of-the-art models.
For sparse implicit feedback scenarios, additional improvement is further achieved by leveraging the step-two implicit feedback information.
\end{abstract}

\begin{IEEEkeywords}
Recommender Systems, Collaborative Filtering, Attentive Model, Neural Networks, Data Mining
\end{IEEEkeywords}

\section{Introduction}
Rapid and accurate prediction for users' preference is the ultimate goal of today's recommender systems \cite{ricci2011introduction}.
Accurate personalized recommender systems benefit both demand-side and supply-side including content publisher and platform\cite{lamere08,googlenews_www07}. For this reason, recommender systems not only attract great interests in the academia \cite{lu2012recommender,rendle2010factorization,hu2008collaborative,juan2016field}, but also are widely developed in industry \cite{koren2009matrix,davidson2010youtube,li2010contextual}.

Matrix factorization (MF) \cite{koren2009matrix} models are one of the remarkable solutions in recommender systems. MF models, as a branch of collaborative filtering techniques, characterize both items and users by vectors in the same space, inferred from the observed entries of the user-item rating matrix; while predicting an unknown rating of a user-item pair relies on the item and the user vectors (the predicted score is usually the inner product of the corresponding item and user vectors). Singular value decomposition (SVD), as a member of MF models, imposes baseline predictors. SVD++ \cite{koren2008factorization} extends SVD by including users' implicit feedback, such as users' historical consumed items, into its model as auxiliary information.
It is found that SVD++ is one of the most effective models in recommender systems.

Besides MF, another successful family of collaborative filtering techniques are the nearest neighbor methods \cite{sarwar2001item,wang2006unifying}, which estimate unknown ratings of a user-item pair by considering similar users' ratings to the item.

Both SVD++ and nearest neighbor models utilize the structure information of user-item bipartite graph. User-item bipartite graph treats each user and item as a vertex. Every rating between users and items is represented as an undirected weighted edge connecting the corresponding user and item vertices. The number of ``step'' on the bipartite graph is defined as the shortest unweighted path length between two vertices. Therefore SVD++, which applies user implicit feedback, utilizes the ``step-one'' structure information whereas nearest neighbor model, which relies on similar users' ratings to predict, makes use of ``step-two'' structure information.

Inspired by the structure of user-item bipartite graph, SVD++ exposes two limitations:
(i). SVD++ only utilizes user-side implicit feedback, whereas item-side implicit feedback is not leveraged;
(ii). SVD++ treats interacted items as users' implicit feedback equally, while in real world a user usually has preference over viewed items. For example, when it comes to movie recommendation, a user has rated two movies: Titanic and Stargate: Continuum.
 In this case, Stargate: Continuum should be much more important because Titanic is watched by so many people, while Stargate: Continuum may only attract Science fiction fans and can better reflect a user's personal interest.

In this paper, we propose three novel models to resolve the above two limitations of SVD++:
\begin{enumerate}
\item Graph-based collaborative filtering (GCF) model generalizes implicit feedback from a graph perspective and introduces item implicit feedback to SVD++ model. As bipartite graph structure information has shown its effectiveness in the aforementioned two branches of CF models, it is natural to consider introducing implicit feedback on items, which describes the relationship between items better and thus improves the performance of the model. 

\item On the basis of GCF model, Weighted Graph-based CF (W-GCF) model applies a matrix form, which models users' personal tastes and items' special popularity, to achieve weight mechanism for the implicit feedback. According to various user tastes and item popularities, different weights are calculated to describe the profiling ability of the implicit feedback to the user and item.

\item Attentive Graph-based CF (A-GCF) model also achieves weight mechanism for implicit feedback. Rather than decomposing weights between the user/item and the implicit feedback, A-GCF model adopts attention network on the implicit feedback to distinguish different importance. Furthermore, the data representation of implicit feedback on a user/item is essentially defined by the ``neighborhood'' on bipartite graph, which can consist of multi-step vertices of user/item rather than step-one vertices in SVD++. In the paper we conduct experiment with step-two implicit feedback data on A-GCF model and prove its effectiveness, especially for sparse data.
\end{enumerate}
In the experiments, we also evaluate and compare the influence of different implicit feedback sampling methods to model performance. We find out that for W-GCF and A-GCF models, random sampling can achieve close performance compared with relevance sampling and greatly saves time for data pretreatment.

The rest of this paper is organized as follows. We discuss the related work in Section \ref{sec:related} and present two preliminary models in \ref{sec:pre}. Then we present our models in Section \ref{sec:model}. Experimental settings and results are discussed in Section \ref{sec:exp}. We finally conclude this paper and discuss the future work in Section \ref{sec:con}.

\section{Related Work} \label{sec:related}
Collaborative filtering (CF) has been well studied for personalized recommender systems during the last decade \cite{schafer2007collaborative,koren2009matrix,rendle2010factorization}. The basic assumption of CF is that the users with similar behavior tend to like the same item, and the items with similar audiences tend to have the same rating \cite{xue2005scalable,wang2006unifying}.

From the model's perspective, CF mainly consists of memory-based and model-based methods \cite{su2009survey}. 
Memory-based methods directly define similarities of pairs of items or pairs of users, based on which preference scores of user-item pairs can be calculated\cite{wang2006unifying,sarwar2001item}.
Despite their simplicity and interpretability, memory based methods usually take too much time or memory to calculate item-item or user-user similarity, making them incapable of handling large-scale recommendation scenarios.
On the other hand, model-based CF methods normally learn the similarities by fitting the model to the user-item interaction data and make the prediction based on the model. Latent factor models are major implementation of model-based methods, such as probabilistic latent semantic analysis (pLSA) \cite{hofmann2004latent} and the most widely used matrix factorization \cite{koren2009matrix}.

Koren proposed SVD++ \cite{koren2008factorization}, a classic model to combine user's ``neighborhood", i.e. previously rated items, and matrix factorization model for prediction, which can be viewed as a mixture of memory-based and model-based methods.

From the data perspective, for the prediction of target data, the classic CF tasks focus on rating prediction, i.e. to make the predicted rating scores as accurate as possible, which is a regression problem \cite{resnick1994grouplens}. Since 2009, there emerge researches on implicit data, where there is only the observation of a user consuming an item (without any explicit rating) \cite{hu2008collaborative}. Such data format can be regarded as a one-class classification problem \cite{manevitz2001one,pan2008one}.

Furthermore, for such implicit feedback data, the recommendation performance is typically evaluated by a top-$N$ recommendation task, i.e. to select and rank $N$ items to the target user and thus the learning to rank evaluation metrics can be adopted, such as mean average precision (MAP), normalized cumulative discounted gain (NDCG), and mean reciprocal ranking (MRR) \cite{baeza1999modern}. We only study the more fundamental regression problem, thus the ranking metrics will not be evaluated on our models, and these ranking approaches will not be compared with our proposal.

As for the side information, such as user demographics, item attributes and the recommendation context, many matrix factorization variants were proposed \cite{koren2009matrix,koren2010collaborative}, among which the most successful model would be factorization machines (FM) \cite{rendle2010factorization}. In FM, the side information and user/item identifiers are regarded as one-hot features of different fields, with a low dimensional latent vector assigned to each feature. The interaction of user, item and side information is formulated by vector inner product.

Recently, with the great success of attention network applied to various applications, such as image recognition \cite{mnih2014recurrent}, natural language generation \cite{sutskever2014sequence}, it is not surprising there emerge some attention models for CF. Attentional factorization machine was leveraged in \cite{xiao2017attentional} where attention network is applied on cross features in factorization machines and assigns different feature combinations with different weights. The author in \cite{loyola2017modeling} proposed an attention-based encoder-decoder architecture for modeling user session and predicting next item on past activities.

Compared with the abundant previous works of CF, this work is positioned on the most classic prediction problem with no side information. Our work makes differences with memory-based and model-based CF methods by considering graph-based user/item representation. Our model is expected to extract more useful information from the simple bipartite graph with the help of attention network, thus can be easily integrated with other feature based frameworks.

\begin{table}[tbp]
    \caption{Notation Table}
    \label{tab:notation}
    \centering
    \begin{tabular}{cl}
        \toprule
        $U$ & User set\\
        $I$ & Item set\\
        $u,v$ & user $u,v \in U$\\
        $i, j$ & item $i,j \in I$\\
        $K$ & dimension of latent factor\\
        $K'$ & dimension of embedding for implicit feedback\\
        $p_u$ & latent factor for user $u$\\
        $q_i$ & latent factor for item $i$\\
        $b_u$ & score bias of user $u$\\
        $b_i$ & score bias of item $i$\\
        $b$ & global score bias\\
        $y_j$ & embedding for user implicit feedback $j$\\
        $x_v$ & embedding for item implicit feedback $v$\\
        $R(u)$ & set of items that has interaction with user $u$\\
        $R(i)$ & set of users that has interaction with item $i$\\
        $N_u$ & number of users\\
        $N_i$ & number of items\\
        $\phi_{uj}$ & weight of user implicit feedback $j$ to user $u$\\
        $\phi_{vi}$ & weight of item implicit feedback $v$ to item $i$\\
        $\alpha_u$ & user $u$ embedding for implicit feedback weight\\
        $\beta_i$ & item $i$ embedding for implicit feedback weight\\
        $R_2(u)$ & set of step-two implicit feedback for user $u$\\
        $R_2(i)$ & set of step-two implicit feedback for item $i$\\
        $a_{uj}$ & attention score of user implicit feedback $j$ to user $u$\\
        $a_{vi}$ & attention score of item implicit feedback $i$ to item $i$\\
        \bottomrule
    \end{tabular}
\end{table}

\section{Preliminary} \label{sec:pre}
In this section, we provide detailed preliminaries of basic matrix factorization and SVD++. In order to study the fundamental CF problems, we choose to focus on (user, item, rating) tuples and the structural information within the user-item bipartite graph. Side information other than user IDs, item IDs, and ratings are not considered in this paper. Such additional features can be seamlessly introduced into our model by leveraging models like factorization machines. All the notations that will show up later in this paper are listed in Table~\ref{tab:notation}.

\subsection{Matrix Factorization}

In basic matrix factorization model, a user $u$ and an item $i$ are represented by latent vectors $p_u$ and $q_i$, respectively, where $p_u,q_i \in \mathbb{R}^K$.
Formally, the rating prediction in basic MF model is formulated as:
\begin{align}
\hat{r}_{ui} = f(\langle p_{u} , q_{i}\rangle + b_{u} + b_{i} + b),
\end{align}
where $u \in \{1, 2, \cdots, N_u\}$, $i \in \{1, 2, \cdots, N_i\}$, $f$ is a non-decreasing scaling function to bound the predicted ratings to certain range, and $\langle p_{u} , q_{i}\rangle$ is the inner product of user and item latent vectors. 

\subsection{SVD++}
Except for a user's identifier, his neighborhood (denoted as $R(u)$, in this case, his rated items), regarded as implicit feedback, also characterizes a user's preference.
Based on this intuition, SVD++ model characterizes a user $u$ by the latent vector $p_u$, as well as those items rated by the user. More formally, the rating prediction of SVD++ model is formulated as:
\begin{align}
\hat{r}_{ui} & = f(\langle p_u + p_{R(u)}, q_i \rangle + b_{u} + b_{i} + b).
\end{align}

In the above formulation, user $u$ is modeled by the latent vector $p_{u}$ and its implicit feedback $p_{R(u)}$:
\begin{align}
p_{R(u)} & = |R(u)|^{-\frac{1}{2}}\sum_{j\in R(u)}y_j \label{equ:pru}
\end{align}
where $y_j \in \mathbb{R}^{K'}$ is independent of item embedding $q_j$, and these items are equally weighted by a constant $|R(u)|^{-\frac{1}{2}}$.
The set of items rated by a user could be recognized as a feature of this user. SVD++ employs this feature in users' representation by taking normalized sum over the rated item set, as shown in \eqref{equ:pru}.

Despite the excellent performance in practice, SVD++ has two limitations as mentioned before. Firstly, SVD++ does not utilize item-side implicit feedback;
secondly, SVD++ treats interacted items equally, which may not hold true in practice.
Thus we propose Weighted Graph-based CF model and Attentive Graph-based CF model.

\section{Methodology}\label{sec:model}

In this section, we propose our Weighted Graph-based CF model and Attentive Graph-based CF model.

\subsection{Weighted Graph-based CF model}

Before introducing weighted version, we firstly introduce Graph-based CF (GCF) model. 
In order to fully utilize item-side graph information, we extend SVD++ by incorporating implicit feedback on items, i.e. the users who have interacted with the given item. For instance, in Netflix dataset, the implicit feedback on a movie is the set of users who have watched and rated it before. In our GCF model, the rating prediction is defined as follows:

\begin{align}
\hat{r}_{ui} & = f(\langle p_u + p_{R(u)}, q_i + q_{R(i)} \rangle + b_{u} + b_{i} + b) \label{equ:cfg}\\
p_{R(u)} & = |R(u)|^{-\frac{1}{2}}\sum_{j\in R(u)}y_j \label{equ:cfg-p}\\
q_{R(i)} & = |R(i)|^{-\frac{1}{2}}\sum_{v \in R(i)}x_v \label{equ:cfg-q}
\end{align}
where $x_v \in \mathbb{R}^{K'}$, and other parameters follow the same definition in SVD++.

We further propose Weighted Graph-based CF (W-GCF) model to tackle the second limitation. Instead of being equally treated as shown in \eqref{equ:cfg-p} and \eqref{equ:cfg-q}, 
the individuals in both users' and items' implicit feedbacks are modeled independently.
In our W-GCF model, \eqref{equ:cfg-p} and \eqref{equ:cfg-q} are re-defined as follows:
\begin{align}
p_{R(u)} & = \sum_{j\in R(u)}\phi_{uj}y_j \label{equ:w-cfg-p}\\
q_{R(i)} & = \sum_{v \in R(i)}\phi_{vi}x_v \label{equ:w-cfg-q}
\end{align}

In \eqref{equ:w-cfg-p} and \eqref{equ:w-cfg-q}, $\phi \in \mathbb{R}^{N_u \times N_i}$ is the weight matrix for every latent factor in the implicit feedback for both users and items. $\phi_{uj}$ describes the profiling capability of user implicit feedback $j$ to user $u$ (similar semantics applies to $\phi_{vi}$). However, if we train $\phi$ directly, we will come across two main obstacles. The first obstacle is the size of $\phi$, which is never feasible in practice. 
The second obstacle is the serious sparsity of $\phi$ as users usually only interact with a small subset of items. 
Thus it is natural to use matrix factorization to learn low rank representation of the weight matrix. We decompose weight matrix into two smaller matrices, $\phi = \alpha \beta^T$, where
$\alpha \in \mathbb{R}^{N_u \times K'}$, $\beta \in \mathbb{R}^{N_i \times K'}$. Such decomposition can also be explained as user's personal taste and item's special popularity respectively. When some items' special popularity meets users' personal taste, such items serve as implicit feedback of the users ought to be assigned a high weight; whereas items' special popularity is not consistent with users' personal taste, then such items serve as implicit feedback have little influence on users and should have a low weight. Therefore in W-GCF model \eqref{equ:w-cfg-p} and \eqref{equ:w-cfg-q} are replaced by: 

\begin{align}
p_{R(u)} & = \sum_{j\in R(u)}(\alpha_{u} \times \beta_{j}^T)y_j \label{equ: weighted-pu}\\
q_{R(i)} & = \sum_{v \in R(i)}(\alpha_{v} \times \beta_{i}^T)x_v \label{equ: weighted-qi}
\end{align}

\subsection{Attentive Graph-based CF model}

\begin{figure}[tbp]
    \centering \includegraphics[width=0.45\textwidth]{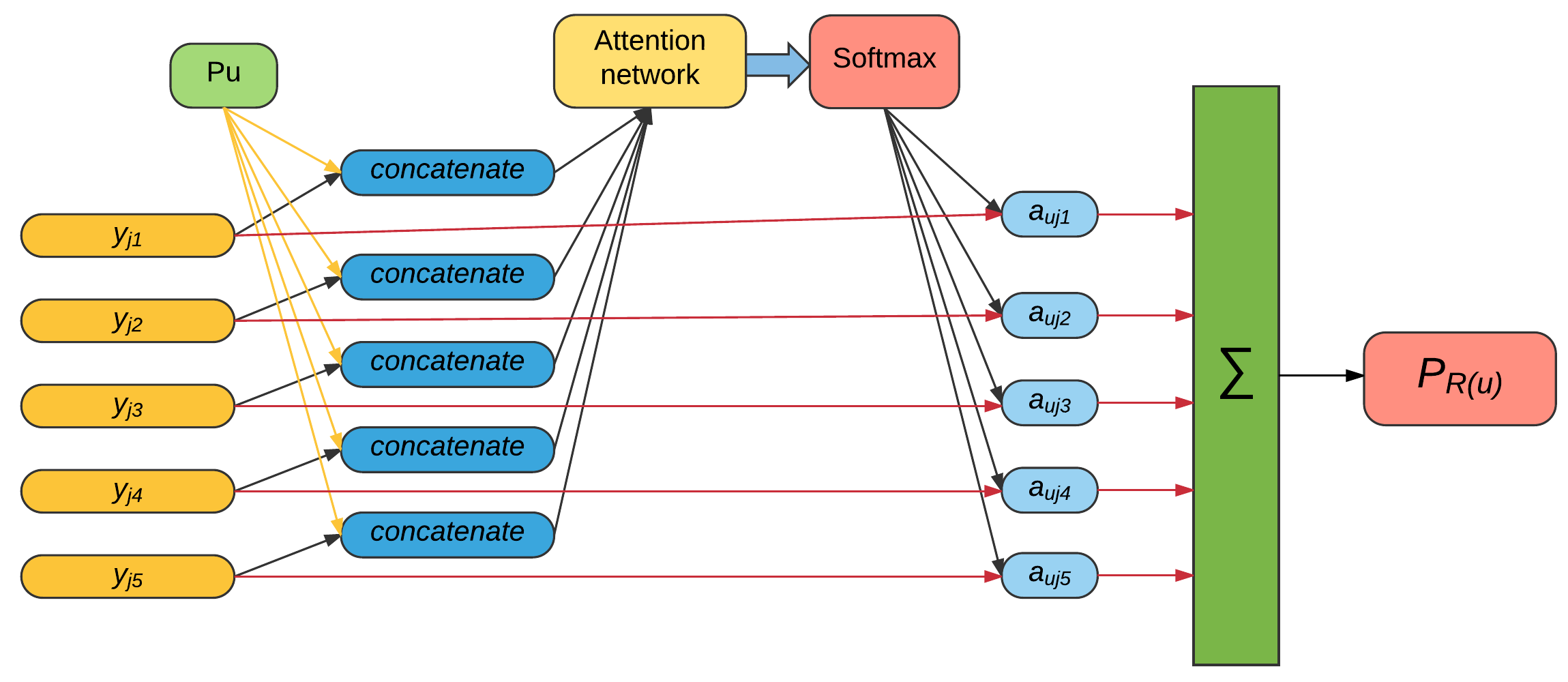}
    \caption{attention network in generation of user implicit feedbacks}\label{fig:attention network}
\end{figure}

Besides matrix representation of user taste and item popularity, another solution is to learn a functional representation, i.e. given user and item hidden vectors, a neural network can be employed to predict the importance of implicit feedback.
From this intuition, we introduce attentive mechanism into our model, in order to automatically learn the importance of each neighbor in a graph.
In Attentive Graph-based CF (A-GCF) model, attention network is applied to substitute the $\alpha$ and $\beta$ matrix in W-GCF model to evaluate the importance between different user/item and their implicit feedbacks. The input of the attention network for user implicit feedback is the concatenation of user embedding and the corresponding implicit feedback embedding. Then the concatenated embeddings go through a multi-layer perceptron (MLP) with ReLU non-linearity. Finally, softmax function is performed over the outcome of MLP to ensure that all attention scores be normalized between $0$ and $1$ with sum of $1$. Formally, the attention network is defined as follows:

\begin{align}
a'_{uj} &= f_{MLP}(p_u, y_j)\\
a_{uj} & = \frac{exp(a'_{uj} / t)}{\Sigma_{j' \in R(u)} exp(a'_{uj'} / t)}
\end{align}
where parameter $t$ denotes the temperature of softmax to properly adjust the variance of the attention scores. 
The same procedure is carried out for item implicit feedback.
Attention scores are then multiplied with implicit feedback embeddings as weights before implicit feedback embeddings are summed together and added in user and item embeddings. \eqref{equ: weighted-pu} and \eqref{equ: weighted-qi} now rewrite as follows:
\begin{align}
p_{R(u)} & = \sum_{j\in R(u)}a_{uj}y_j \label{equ:a-cfg-p}\\
q_{R(i)} & = \sum_{v\in R(i)}a_{vi}x_v \label{equ:a-cfg-q}
\end{align}
where $a_{uj}$ and $a_{vi}$ are the attention score for user implicit feedbacks and item implicit feedbacks separately, just as $\phi_{uj}$ and $\phi_{vi}$ in W-GCF model. Fig.~\ref{fig:attention network} shows one example of how user implicit feedback $p_{R(u)}$ is generated.

\subsection{Feedback Process}\label{subsec:feedback-sampling}
\begin{figure}[t]
    \centering \includegraphics[width=0.45\textwidth]{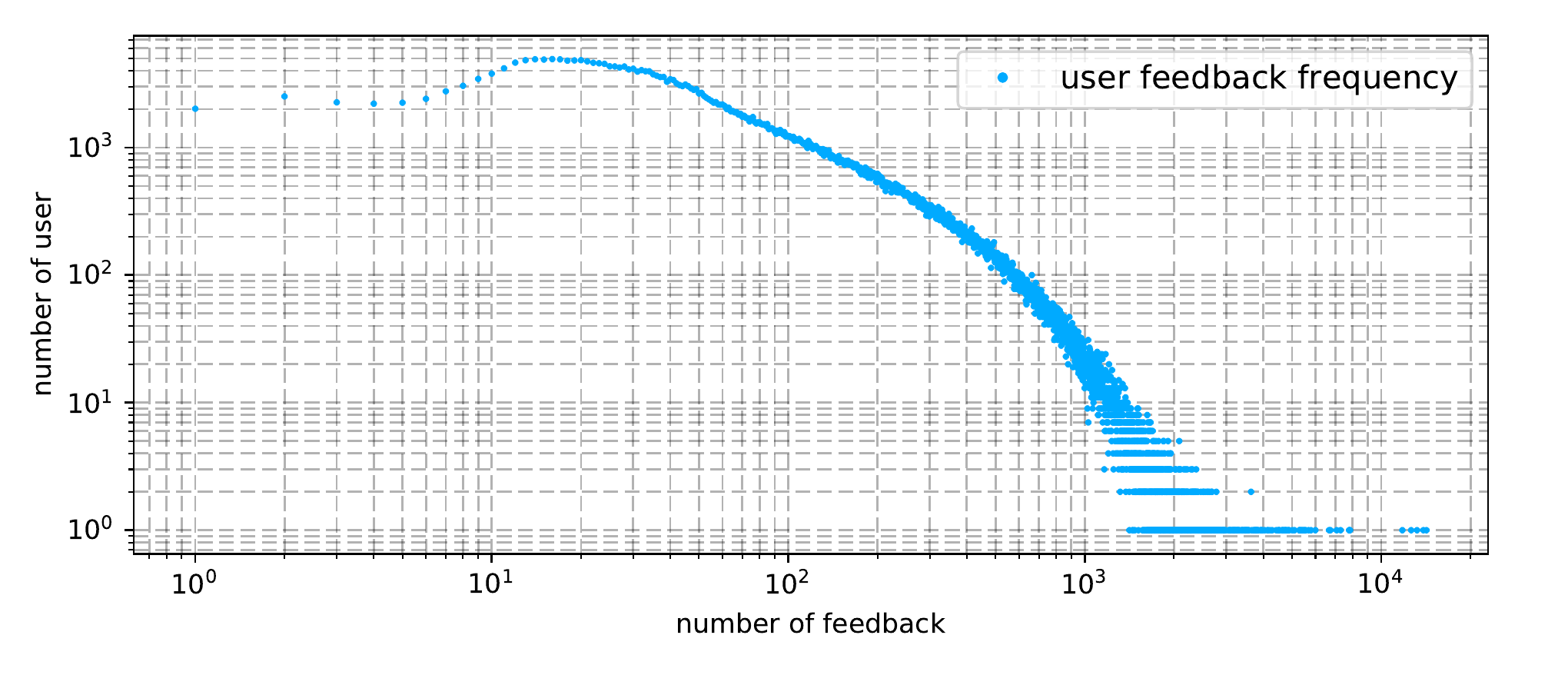}
    \centering \includegraphics[width=0.45\textwidth]{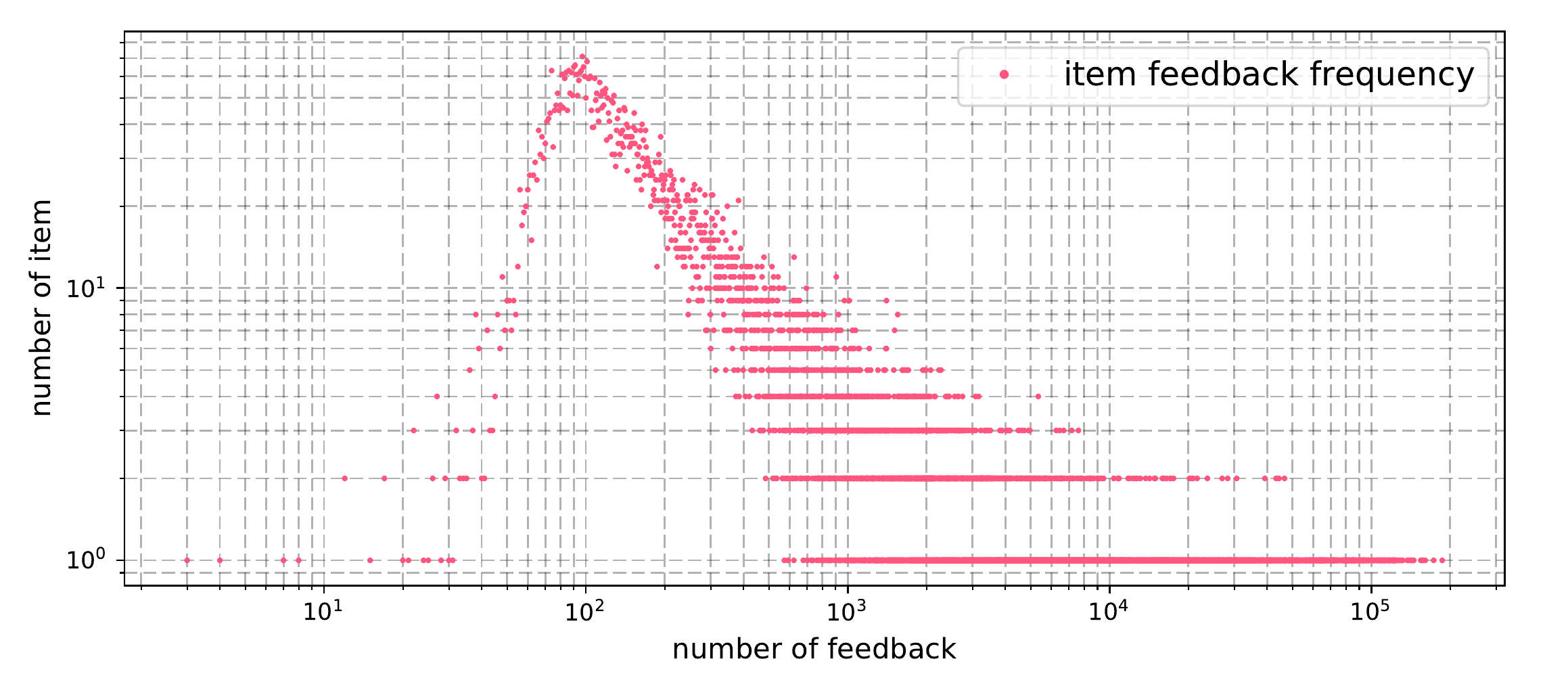}
    \caption{User and Item Feedback Distribution on NetFlix}\label{fig:3}
\end{figure}
In practice, the number of implicit feedback, i.e. the number of interacted users, on popular items in datasets is usually large.
We depict the statistics for Netflix dataset in Fig.~\ref{fig:3}. X-axis of Fig.~\ref{fig:3} is the number of implicit feedback (i.e. interacted items or interacted users) and Y-axis represents for how many users or items have a specific number of implicit feedback. Every point on the figures represents one class of users or items who have the same number of feedbacks. As can be observed, the number of implicit feedbacks on both users and items are distributed in long-tail.

We can tell from Fig.~\ref{fig:3} that most users have $10^1$ to $10^2$ feedbacks and the number of feedbacks for items is most likely between $10^2$ and $10^3$. However, there still exist items which have more than $10^5$ feedbacks because such popular items interact with many users. Obviously, it is not practical to include all the feedback data within a model, especially for W-GCF and A-GCF model, where every implicit feedback needs to multiply its weight or attention score first. Therefore, sampling is quite necessary to include implicit feedback in our models.

In the experiments we study two sampling policies, which are ``relevance'' and ``random''. For ``relevance'' sampling method, firstly, we run a basic matrix factorization to generate embeddings for users and items. Secondly, we select the relevant items for each user, where the ``relevance" of a user and an item is measured by inner-product of the corresponding embedding vectors. Top-20 most relevant items are selected for each user as the implicit feedback in the model. Thirdly, the same sampling procedure is also performed to generate implicit feedback for each item. For users who rate less than 20 items, we take all the items as implicit feedback and pad a unified and fake item repeatedly until the size of implicit feedback of this user reaches 20. Exactly the same trick is applied to the items which are rated by less than 20 users. For ``random'' sampling method, user and item implicit feedbacks are taken randomly from all candidates with replacement. So for users who have rated less than 20 items or items with less 20 ratings, duplicated items/users will appear in their implicit feedbacks.

Except for step-one implicit feedback, we also sample step-two implicit feedbacks for each user and item to further figure out how graph structure helps to predict scores of unknown user-item links on bipartite graph. For step-two implicit feedbacks of user $u$, firstly we generate candidates set $\bigcup\limits_{i_j \in R(u)} R(i_j)$, which is the union of item implicit feedbacks for all items in $u$'s implicit feedback list, either by ``relevance'' or ``random'' method. Then 20 step-two implicit feedbacks for user $u$, denoted as $R_2(u)$ are randomly sampled from the candidate set. There are always more than 20 candidates so no padding is needed here. The same trick is applied to sample item two step implicit feedbacks $R_2(i)$. 

\section{Experiments}\label{sec:exp}
\subsection{Dataset}

Our experiments are conducted on the original Netflix Prize contest. Netflix dataset collects in total $100,467,535$ records from $17,770$ users over $479,837$ movies.
Each record contains a rating score from $1$ to $5$ where $5$ indicates user prefers the item most.
In our experiments, all scores are normalized into range $[0,1]$.
We use 80\% of the dataset as train set and the rest 20\% as test set.
Our dataset splitting strategy ensures that all users and movies in test set appear at least once in training set.

\subsection{Comparing Models}
During experiments, we compare the following models:

\begin{table*}
    \caption{Evaluations on NetFlix Dataset}
    \label{table:rmse}
    \centering
    \begin{tabular}{rcc|rcc}
        \toprule
        Model& random sampling & relevance sampling & Model & random sampling & relevance sampling\\
        \midrule
        MF & 0.173405 & - & & &\\
        NCF & 0.173266 & - & & &\\
        NFM & 0.169594 & - & & &\\
        SVD++& 0.172743 & 0.171890 & GCF & 0.17101 & 0.17062\\
        W-SVD++ & 0.169570 & 0.169293 & W-GCF& \textbf{0.168350} & 0.168639\\
        A-SVD++ & 0.168505 & 0.168425 & A-GCF& 0.168426 & 0.168814\\
        \bottomrule
    \end{tabular}
\end{table*}

\begin{itemize}
    \item Matrix Factorization (\textbf{MF}) model. Both users and items are characterized by vectors in the same space, inferred from the historical user-item interactions. The predicted score of a user-item pair is generated by the inner product of corresponding user's and item's vectors.
    \item \textbf{SVD++} model. This model extends MF and further leverages users' historical interacted items as users' implicit feedbacks.
    \item Graph-based collaborative filtering (\textbf{GCF}) model. GCF model extends SVD++ model by introducing the item implicit feedbacks (list of users who have historical interaction with the item) into the model.
    \item Weighted Graph-based CF (\textbf{W-GCF}) model. Instead of setting an equal and fixed weight for individuals in implicit feedbacks as in SVD++ and GCF, W-GCF further learns the weights of individuals in implicit feedbacks in a matrix form.
    \item Attentive Graph-based CF (\textbf{A-GCF}) model. Compared with W-GCF model, A-GCF model applies attention network to distinguish the importance of implicit feedbacks.
    \item Weighted SVD++ (\textbf{W-SVD++}) model. This model introduces dynamic weights for user implicit feedback for SVD++ model.
    \item Attentive SVD++ (\textbf{A-SVD++}) model. This model applies attention network for user implicit feedback for SVD++ model.
\end{itemize}
To compare our models with the most recent technology, we also conduct experiments on following models:
\begin{itemize}
    \item Neural Collaborative Filtering (\textbf{NCF}) model. This model combines matrix factorization and multi-layer perception (MLP) under one neural network. In NCF model, embeddings of MF and MLP are independent. The output of MF here is the element-wise product of the embeddings to maintain the same vector form as MLP. The outputs of MF and MLP are then combined together with a hyper-parameter to determine the trade-off between MF and MLP.
    \item Neural Factorization Machine (\textbf{NFM}) model. This model applies neural network on output of traditional factorization machine to introduce nonlinear component. In NFM model, the corresponding embeddings of second-order feature interaction of factorization machine first do an element-wise product and then go through multi-layer perception. Finally, the output of MLP gives the prediction together with global bias and first-order linear regression of traditional factorization machine.
\end{itemize}

All the models are implemented with TensorFlow. We rewrite NCF model and NFM model according to \cite{he2017neural} and \cite{he2017neuralfm} respectively. For NCF model we add extra bias to the MF part to fit the task. We share the code for repeatable experiments\footnote{The experiment code: https://goo.gl/kcRH3D. We will publish the code on GitHub upon the paper acceptance.}.

\subsection{Evaluation Matrics}

In our experiment, we adopt the evaluation matrix of root mean square error ($\mathbf{RMSE}$).
This metric is widely used for score-based recommender systems, which is also used in Netflix Prize contest.

\subsection{Performance}

In the experiments, we compare our GCF, W-GCF, and A-GCF models to MF, SVD++ and NCF, NFM model in setting where $K=16$ with respect to RMSE. We set the dimension of latent vectors of implicit feedbacks $K' = K$ in the experiments. The comparison results are presented in Table~\ref{table:rmse}. Note that MF, NFM, and NCF have no implicit feedback so sampling policy and item implicit feedback are not applicable.

\subsubsection{Overall performance} From Table~\ref{table:rmse}, we can observe that all models outperform MF, of which the performance is $0.1734$. Among all models best performance $0.168350$ is achieved by W-GCF model with random sampling, gaining $2.54\%$ improvement over SVD++ model. Comparing with NFM and NCF models, W-GCF also gains $0.73\%$ improvement over NFM model. NCF model does not perform well on RMSE task maybe because it does not adopt $L_{2}$ regularization.

\subsubsection{Item implicit feedback effectiveness} Table~\ref{table:rmse} is divided into two sides. Models like W-SVD++ on the left utilizes only user side implicit feedback information whereas models like W-GCF on the right side apply both user and item implicit feedbacks. Though MF, NFM, and NCF utilize no implicit feedback, here we still categorize them to the left side. Comparing the performance of models in two sides of the table, we find out that all the three models in the right side outperform the corresponding models on the left side, except for A-GCF model with relevance sampling. In addition, GCF and W-GCF gain performance improvement of $0.72\% - 0.74\%$ whereas item implicit feedback shows a negligible effect on A-GCF model, which may be caused by potential overfitting problem.

\subsubsection{Sampling policy} In our experiments, for all the models except for MF, users' and items' implicit feedbacks are restricted to a fixed number (namely, 20) by sampling or padding. In the experiments, we take two sampling methods, ``random'' and ``relevance'', as discussed before in subsection \ref{subsec:feedback-sampling}.

For SVD++ model, relevance sampling method gets the performance of $0.17189$ while random sampling method only gets $0.172743$. However, when it comes to W-GCF and A-GCF, things become different. When comparing two sampling methods in the same rows of Table~\ref{table:rmse}, the performance gap between random and relevance sampling becomes minus for A-GCF and W-GCF models. This is probably caused by the ability of attention network or the weight matrix to distinguish important implicit feedbacks from irrelevant ones. Especially for A-GCF model, there is no significant necessity to sample implicit feedbacks according to their relevance to users or items and therefore it will save time for the training process. Another potential benefit is that it also allows the possibility for W-GCF and A-GCF model to adapt for stream data.

\subsection{Effectiveness of step-two implicit feedback}
\begin{figure}[tbp]
    \centering \includegraphics[width=0.45\textwidth]{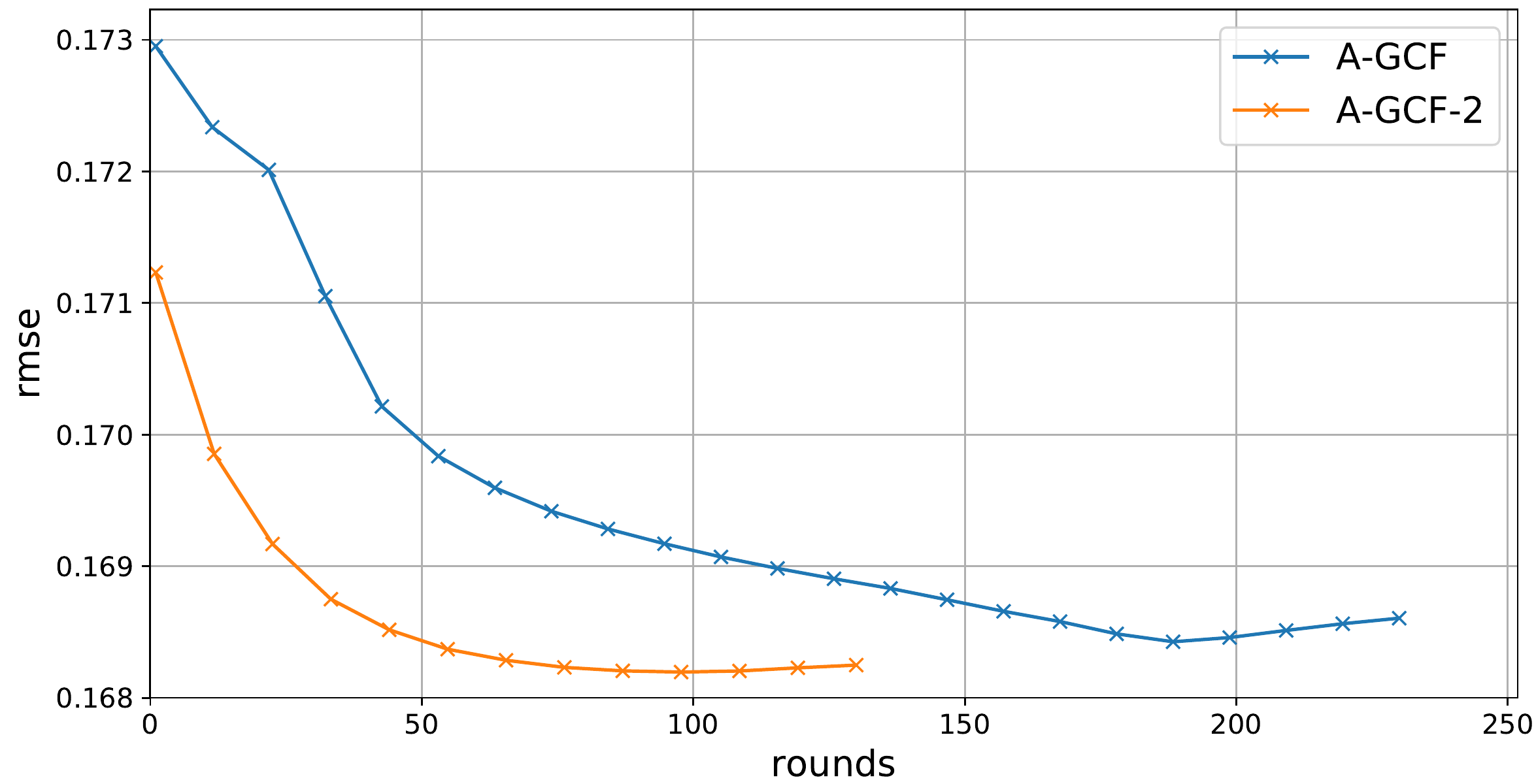}
    \caption{learning curve of A-GCF and A-GCF-2 models} \label{fig:step_two_curve}
\end{figure}

Besides step-one implicit feedbacks, we also run experiments over step-two implicit feedbacks on A-GCF, which is called A-GCF-2 model. We treat step-two implicit feedbacks the same way as step-one implicit feedbacks except that they do not share embeddings. Detailed experiment result is shown in Fig.~\ref{fig:step_two_curve}. The learning curve from Fig.~\ref{fig:step_two_curve} demonstrates that adding step-two implicit feedbacks of user and item not only slightly improves the overall RMSE performance from $0.168426$ to $0.168196$ over test data, but also speeds up the convergence of the model. With the same setting of learning parameters, A-GCF-2 model takes less than a half of number of rounds to converge than with only step-one implicit feedbacks.

\begin{figure}[tbp]
    \centering \includegraphics[width=0.45\textwidth]{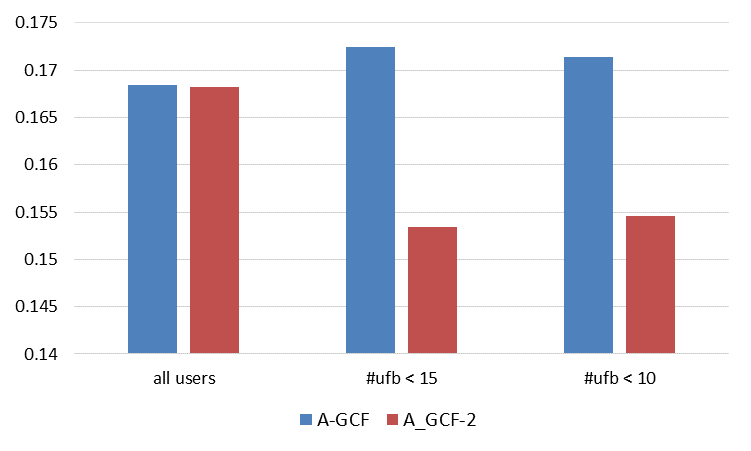}
    \caption{Sparse implicit feedback on A-GCF and A-GCF-2 models} \label{fig:step-two}
\end{figure}

In order to further explore the effect of step-two implicit feedbacks, we also check the performance of A-GCF and A-GCF-2 models on those users who have less step-one implicit feedbacks. In the experiment, we filter the records whose user has less than $10$ and $15$ step-one implicit feedbacks respectively and then calculate RMSE for these records on A-GCF and A-GCF-2 models. The result is shown in Fig.~\ref{fig:step-two}. Through Fig.~\ref{fig:step-two} we can find out that A-GCF-2 model even gets more accurate predictions on sparse implicit feedbacks while A-GCF model, on the contrary, performs worse than A-GCF model with full dataset. Though two models performs closely with full dataset, A-GCF-2 model outperforms A-GCF model by $9.83\% - 11.05\%$ when the number of user feedback is less than $15$ and $10$. Obviously, step-two implicit feedback provides useful additional information about user and item for A-GCF-2 model.

\subsection{Parameters}
\begin{figure}[htbp]
    \centering \includegraphics[width=0.45\textwidth]{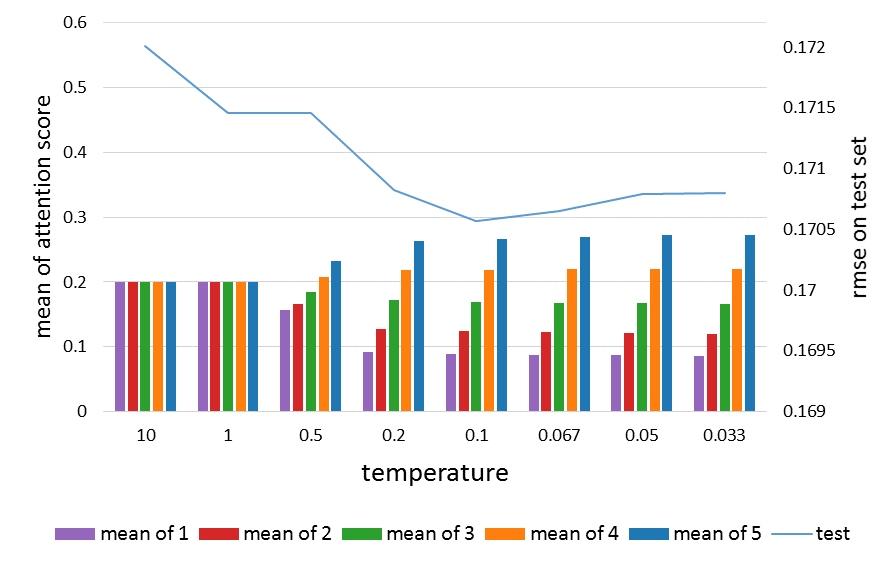}
    \caption{A-GCF model under different temperature settings}\label{fig:6}
\end{figure}
Compared with traditional SVD++, W-GCF and A-GCF models both introduce weight mechanism as well as item side implicit feedback. A-GCF applies attention network and W-GCF adopts matrix form. The following two subsections give a detailed discussion about parameter tuning for the attention network in A-GCF model and the weight matrix in W-GCF model. 

\textbf{Temperature in attention network} In this section, we compare the performance of A-GCF model under different softmax temperature parameters. All the experiments are conducted with same parameters and training rounds except for the softmax temperature $t$ and the result is shown in Fig.~\ref{fig:6}. In order to show the ability of attention network to distinguish important implicit feedbacks, for each pair of user/item and its implicit feedback, we find the original rating (before normalization) through the training data and categorize all the pairs according to the rating, namely $1, 2, \dots, 5$, into $5$ groups. Then we calculate the corresponding mean attention score for each group, which in Fig.~\ref{fig:6} is mean of $1,2,\dots,5$, respectively. Finally, we compare the mean attention scores with the original rating which is served as ground truth here. As a high-rating item should be more important to user than a low-rating item, attention scores should reflect this and have high values on high-rating items. This conclusion also holds for item implicit feedbacks and corresponding attention scores.

From Fig.~\ref{fig:6} we can see that A-GCF model with temperature of $10$ gets a relatively high RMSE performance on test set, which is obvious as a large temperature value restricts the final attention score merely near the average value. As the temperature gets smaller, the RMSE value on test after 30 rounds' training continues to drop until temperature $t = 0.1$ and then it increases again. The reason is that as the temperature decreases, attention scores start to take their part as weights to emphasize those ``important'' implicit feedback embeddings and therefore generate more accurate predictions. However, smaller temperature also enlarges the perturbation during random initialization, which results for the bad performance for model with temperature smaller than $0.05$. Therefore after this series of experiment, we take temperature $t = 0.1$, which is a proper value from Fig.~\ref{fig:6}, for all attentive models.

\begin{figure}[tbp]
    \centering \includegraphics[width=0.45\textwidth]{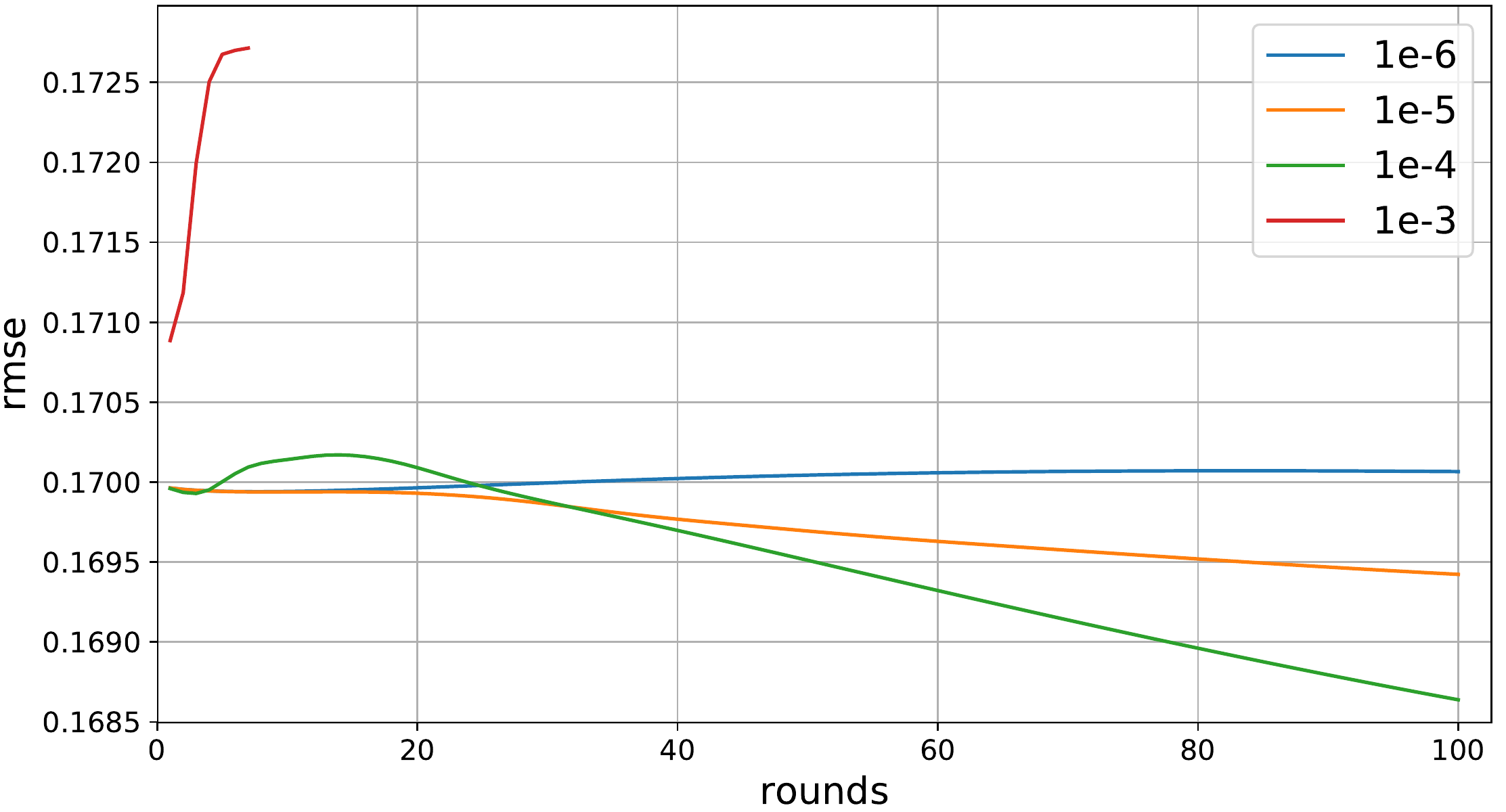}
    \caption{learning curve of W-GCF model under different regularization strength}\label{fig:regularization}
\end{figure}

\textbf{Regularization} In order to prevent overfitting, during model training we apply $L_{2}$ regularization to embedding $p_u, q_i, y_j$ for all models except for NFM and NCF models ($L_{2}$ is only applied on $p_u$ and $q_i$ for MF). For GCF, A-GCF, and W-GCF models, $L_{2}$ regularization is also applied on $x_v$. Except for that, we apply $L_{2}$ regularization specifically to weights $\alpha$ and $\beta$ for W-GCF model.
Fig.~\ref{fig:regularization} shows the learning curve of W-GCF model under different weights regularization coefficients $\lambda_{r}$.

All experiments are conducted under the same parameters for $100$ rounds. Through a series of experiments, we found out that when $\lambda_{r}$ is too small like $10^{-6}$, the curve overfits quickly; while $\lambda_{r}$ is too large as for $\lambda_{r} = 10^{-3}$, $L_{2}$ penalty accounts for most of the loss function and the training process goes wrong and gives back NAN after $10$ rounds or so. Only with $\lambda_{r} = 10^{-4}$, W-GCF model gets the balance between optimizing the model and penalty for model complexity and gets the best performance.

\section{Conclusion and Future Work}\label{sec:con}

In this paper, we study the task of leveraging implicit feedback in graph-based collaborative filtering for recommender systems.
Unlike existing works, we generalize the implicit feedback used in collaborative filtering models to further incorporate both users and items as their particular neighborhood information in the user-item bipartite graph.
We also extend the model by employing adaptive weighting over the implicit feedbacks.
Further, we perform weight matrices and attention networks for each implicit feedback built based on the learned user/item representations.
Extended neighborhood in the bipartite graph such as step-two information can also be utilized in such attentive model to improve performance on sparse data.
Experiments on a well-known collaborative filtering benchmark dataset demonstrate the superiority of our proposed model over the state-of-the-art ones in the classic rating prediction task, especially for the sparse implicit feedback scenarios.

For future work, we will extend our model to other real-world application scenarios, such as top-N item recommendation, where the ranking objective and negative item sampling strategies will be adopted.
Besides, we may also explore how to extend the implicit feedbacks to leverage other structures besides step-two information.

\bibliographystyle{IEEEtran}
\bibliography{cfg-net}

\begin{thebibliography}{10}
\providecommand{\url}[1]{#1}
\csname url@samestyle\endcsname
\providecommand{\newblock}{\relax}
\providecommand{\bibinfo}[2]{#2}
\providecommand{\BIBentrySTDinterwordspacing}{\spaceskip=0pt\relax}
\providecommand{\BIBentryALTinterwordstretchfactor}{4}
\providecommand{\BIBentryALTinterwordspacing}{\spaceskip=\fontdimen2\font plus
\BIBentryALTinterwordstretchfactor\fontdimen3\font minus
  \fontdimen4\font\relax}
\providecommand{\BIBforeignlanguage}[2]{{%
\expandafter\ifx\csname l@#1\endcsname\relax
\typeout{** WARNING: IEEEtran.bst: No hyphenation pattern has been}%
\typeout{** loaded for the language `#1'. Using the pattern for}%
\typeout{** the default language instead.}%
\else
\language=\csname l@#1\endcsname
\fi
#2}}
\providecommand{\BIBdecl}{\relax}
\BIBdecl

\bibitem{ricci2011introduction}
F.~Ricci, L.~Rokach, and B.~Shapira, \emph{Introduction to recommender systems
  handbook}.\hskip 1em plus 0.5em minus 0.4em\relax Springer, 2011.

\bibitem{lamere08}
P.~Lamere and S.~Green., ``Project aura: recommendation for the rest of us,''
  \emph{Presentation at Sun JavaOne Conference}, 2008.

\bibitem{googlenews_www07}
A.~S. Das, M.~Datar, A.~Garg, and S.~Rajaram, ``Google news personalization:
  scalable online collaborative filtering,'' in \emph{WWW}, 2007.

\bibitem{lu2012recommender}
L.~L{\"u}, M.~Medo, C.~H. Yeung, Y.-C. Zhang, Z.-K. Zhang, and T.~Zhou,
  ``Recommender systems,'' \emph{Physics Reports}, vol. 519, no.~1, pp. 1--49,
  2012.

\bibitem{rendle2010factorization}
S.~Rendle, ``Factorization machines,'' in \emph{Data Mining (ICDM), 2010 IEEE
  10th International Conference on}.\hskip 1em plus 0.5em minus 0.4em\relax
  IEEE, 2010, pp. 995--1000.

\bibitem{hu2008collaborative}
Y.~Hu, Y.~Koren, and C.~Volinsky, ``Collaborative filtering for implicit
  feedback datasets,'' in \emph{Data Mining, 2008. ICDM'08. Eighth IEEE
  International Conference on}.\hskip 1em plus 0.5em minus 0.4em\relax Ieee,
  2008, pp. 263--272.

\bibitem{juan2016field}
Y.~Juan, Y.~Zhuang, W.-S. Chin, and C.-J. Lin, ``Field-aware factorization
  machines for ctr prediction,'' in \emph{Proceedings of the 10th ACM
  Conference on Recommender Systems}.\hskip 1em plus 0.5em minus 0.4em\relax
  ACM, 2016, pp. 43--50.

\bibitem{koren2009matrix}
Y.~Koren, R.~Bell, and C.~Volinsky, ``Matrix factorization techniques for
  recommender systems,'' \emph{Computer}, vol.~42, no.~8, 2009.

\bibitem{davidson2010youtube}
J.~Davidson, B.~Liebald, J.~Liu, P.~Nandy, T.~Van~Vleet, U.~Gargi, S.~Gupta,
  Y.~He, M.~Lambert, B.~Livingston \emph{et~al.}, ``The youtube video
  recommendation system,'' in \emph{Proceedings of the fourth ACM conference on
  Recommender systems}.\hskip 1em plus 0.5em minus 0.4em\relax ACM, 2010, pp.
  293--296.

\bibitem{li2010contextual}
L.~Li, W.~Chu, J.~Langford, and R.~E. Schapire, ``A contextual-bandit approach
  to personalized news article recommendation,'' in \emph{Proceedings of the
  19th international conference on World wide web}.\hskip 1em plus 0.5em minus
  0.4em\relax ACM, 2010, pp. 661--670.

\bibitem{koren2008factorization}
Y.~Koren, ``Factorization meets the neighborhood: a multifaceted collaborative
  filtering model,'' in \emph{Proceedings of the 14th ACM SIGKDD international
  conference on Knowledge discovery and data mining}.\hskip 1em plus 0.5em
  minus 0.4em\relax ACM, 2008, pp. 426--434.

\bibitem{sarwar2001item}
B.~Sarwar, G.~Karypis, J.~Konstan, and J.~Riedl, ``Item-based collaborative
  filtering recommendation algorithms,'' in \emph{Proceedings of the 10th
  international conference on World Wide Web}.\hskip 1em plus 0.5em minus
  0.4em\relax ACM, 2001, pp. 285--295.

\bibitem{wang2006unifying}
J.~Wang, A.~P. De~Vries, and M.~J. Reinders, ``Unifying user-based and
  item-based collaborative filtering approaches by similarity fusion,'' in
  \emph{Proceedings of the 29th annual international ACM SIGIR conference on
  Research and development in information retrieval}.\hskip 1em plus 0.5em
  minus 0.4em\relax ACM, 2006, pp. 501--508.

\bibitem{schafer2007collaborative}
J.~Schafer, D.~Frankowski, J.~Herlocker, and S.~Sen, ``Collaborative filtering
  recommender systems,'' \emph{The adaptive web}, pp. 291--324, 2007.

\bibitem{xue2005scalable}
G.-R. Xue, C.~Lin, Q.~Yang, W.~Xi, H.-J. Zeng, Y.~Yu, and Z.~Chen, ``Scalable
  collaborative filtering using cluster-based smoothing,'' in \emph{Proceedings
  of the 28th annual international ACM SIGIR conference on Research and
  development in information retrieval}.\hskip 1em plus 0.5em minus 0.4em\relax
  ACM, 2005, pp. 114--121.

\bibitem{su2009survey}
X.~Su and T.~M. Khoshgoftaar, ``A survey of collaborative filtering
  techniques,'' \emph{Advances in artificial intelligence}, vol. 2009, p.~4,
  2009.

\bibitem{hofmann2004latent}
T.~Hofmann, ``Latent semantic models for collaborative filtering,'' \emph{ACM
  Transactions on Information Systems (TOIS)}, vol.~22, no.~1, pp. 89--115,
  2004.

\bibitem{resnick1994grouplens}
P.~Resnick, N.~Iacovou, M.~Suchak, P.~Bergstrom, and J.~Riedl, ``Grouplens: an
  open architecture for collaborative filtering of netnews,'' in
  \emph{Proceedings of the 1994 ACM conference on Computer supported
  cooperative work}.\hskip 1em plus 0.5em minus 0.4em\relax ACM, 1994, pp.
  175--186.

\bibitem{manevitz2001one}
L.~M. Manevitz and M.~Yousef, ``One-class svms for document classification,''
  \emph{Journal of Machine Learning Research}, vol.~2, no. Dec, pp. 139--154,
  2001.

\bibitem{pan2008one}
R.~Pan, Y.~Zhou, B.~Cao, N.~N. Liu, R.~Lukose, M.~Scholz, and Q.~Yang,
  ``One-class collaborative filtering,'' in \emph{Data Mining, 2008. ICDM'08.
  Eighth IEEE International Conference on}.\hskip 1em plus 0.5em minus
  0.4em\relax IEEE, 2008, pp. 502--511.

\bibitem{baeza1999modern}
R.~Baeza-Yates, B.~Ribeiro-Neto \emph{et~al.}, \emph{Modern information
  retrieval}.\hskip 1em plus 0.5em minus 0.4em\relax ACM press New York, 1999,
  vol. 463.

\bibitem{koren2010collaborative}
Y.~Koren, ``Collaborative filtering with temporal dynamics,''
  \emph{Communications of the ACM}, vol.~53, no.~4, pp. 89--97, 2010.

\bibitem{mnih2014recurrent}
V.~Mnih, N.~Heess, A.~Graves \emph{et~al.}, ``Recurrent models of visual
  attention,'' in \emph{Advances in neural information processing systems},
  2014, pp. 2204--2212.

\bibitem{sutskever2014sequence}
I.~Sutskever, O.~Vinyals, and Q.~V. Le, ``Sequence to sequence learning with
  neural networks,'' in \emph{Advances in neural information processing
  systems}, 2014, pp. 3104--3112.

\bibitem{xiao2017attentional}
J.~Xiao, H.~Ye, X.~He, H.~Zhang, F.~Wu, and T.-S. Chua, ``Attentional
  factorization machines: Learning the weight of feature interactions via
  attention networks,'' \emph{arXiv preprint arXiv:1708.04617}, 2017.

\bibitem{loyola2017modeling}
P.~Loyola, C.~Liu, and Y.~Hirate, ``Modeling user session and intent with an
  attention-based encoder-decoder architecture,'' in \emph{Proceedings of the
  Eleventh ACM Conference on Recommender Systems}.\hskip 1em plus 0.5em minus
  0.4em\relax ACM, 2017, pp. 147--151.

\bibitem{he2017neural}
X.~He, L.~Liao, H.~Zhang, L.~Nie, X.~Hu, and T.-S. Chua, ``Neural collaborative
  filtering,'' in \emph{Proceedings of the 26th International Conference on
  World Wide Web}.\hskip 1em plus 0.5em minus 0.4em\relax International World
  Wide Web Conferences Steering Committee, 2017, pp. 173--182.

\bibitem{he2017neuralfm}
X.~He and T.-S. Chua, ``Neural factorization machines for sparse predictive
  analytics,'' \emph{SIGIR}, 2017.

\end{thebibliography}

\end{document}